%% file: ifacconf_sim2realxDT.tex
\documentclass{ifacconf}

\usepackage{graphicx}      
\usepackage{natbib}        
\usepackage{amsmath} 
\usepackage{amsfonts} 
\usepackage{algorithm}
\usepackage{algpseudocode}
\usepackage{subfig} 

\usepackage{scalefnt}
\usepackage{tikzscale}
\usepackage{tikz}
\usepackage{pgfplots}
\usepackage{changepage}
\usetikzlibrary{shapes.geometric}
\usetikzlibrary{calc}
\usetikzlibrary{positioning}
\usetikzlibrary{intersections}

\usepackage{fancyhdr}
\fancypagestyle{firstpage}
{
	\fancyhead[L]{}    
	\fancyhead[R]{ © 2022. This work has been accepted to IFAC for publication under a Creative Commons Licence CC-BY-NC-ND}
}
\begin{document}

\begin{frontmatter}

\title{Sim2real for Autonomous Vehicle Control using Executable Digital Twin} 


\author[1,2]{Jean Pierre Allamaa} 
\author[2]{Panagiotis Patrinos} 
\author[1]{Herman Van der Auweraer}
\author[1]{Tong Duy Son}


\address[1]{Siemens Digital Industries Software,  3001, Leuven, Belgium\\(e-mail: jean.pierre.allamaa@siemens.com)}
\address[2]{Dept. Electr. Eng. (ESAT) - STADIUS research group, KU Leuven, 3001 Leuven, Belgium}

\begin{abstract} 
In this work, we propose a sim2real method to transfer and adapt a nonlinear model predictive controller (NMPC) from simulation to the real target system based on executable digital twin (xDT). The xDT model is a high fidelity vehicle dynamics simulator, executable online in the control parameter randomization and learning process. The parameters are adapted to gradually improve control performance and deal with changing real-world environment. In particular, the performance metric is not required to be differentiable nor analytical with respect to the control parameters and system dynamics are not necessary linearized. Eventually, the proposed sim2real framework leverages altogether online high fidelity simulator, data-driven estimations, and simulation based optimization to transfer and adapt efficiently a controller developed in simulation environment to the real platform. Our experiment demonstrates that a high control performance is achieved without tedious time and labor consuming tuning.
\end{abstract}


\end{frontmatter}
\thispagestyle{firstpage}
\pagenumbering{gobble}
\section{Introduction}

With the advancement of control and planning algorithms for autonomous driving, a challenge still remains in the transfer from simulation to the real world. Manual tuning to validate the design requirements in a physical environment with various scenarios is time-consuming and expensive. The automotive industry is trying to leverage more simulation to reduce the physical tuning efforts. Digital Twin (DT) of a vehicle is an accurate and reliable representation model in a complex high degrees of freedom simulation system that supports the vehicle control algorithm development and validation processes (\cite{VanDerAuweraerH2018Sadt}). DTs allow us to test the autonomous driving (AD) algorithms in several configurations, traffic scenarios, and edge cases. Nevertheless, this often comes at the expense of over tuning in simulation hindering transferability to the real world. In recent advancements of control strategies for safe AD, based on reinforcement learning, neural networks, or non-linear model predictive control, transferring and embedding the development from simulation to real systems is a bottleneck. Moreover, no simulator guarantees perfect reliability given a reality gap (sim2real gap) due to erroneous sensor readings, observability of the system, actuation and process noise, and importantly an external environment different from simulation to reality.

Research from the robotics and reinforcement learning communities is currently tackling the problem of transferability. Under the umbrella of transfer learning, three main methods deal with sim2real: domain randomization, domain adaptation, and high-fidelity simulation. In \cite{DBLP:conf/corl/MuellerDGK18}, end-to-end driving policies are transferred from simulation to reality via modularity and abstraction: the learning process becomes indifferent to the environment. \cite{DBLP:journals/pami/MuratoreG021} discuss the presence of a simulation optimization bias (SOB) that overestimates the maximum expected return in a Markovian decision process trained in simulation, in comparison with the real performance. The algorithm is then trained to compensate for the SOB by randomizing the environment. Relying more on high-fidelity simulation, \cite{DBLP:journals/corr/abs-1912-06321} run parallel tests in simulation and real-world while varying the simulation parameters. The authors then measure the success rate in both worlds to find the parameters correlating both domains the most. Eventually, the reliability of the simulator is evaluated using a sim-vs-real correlation coefficient (SRCC). In general, most of the presented methods rely on a large amount of data and scenarios to train the algorithm. Aiming more towards online adaptation, \cite {DBLP:journals/corr/abs-2012-05841} show the importance of a reliable predictive model in the shape of a DT incorporating real-world data. They use a graphical model method to describe the evolution of the DT dynamical systems. The DT is integrated for dynamic decision making in a predictive process to monitor the health and safety of the unmanned aerial vehicle.

The main contribution in this work is to suggest a framework based on an executable digital twin (xDT), combining randomization and adaptation with real data to transfer a control strategy from sim2real. The xDT is fit for control purposes and can be implemented in a vehicle ECU as it is a self-contained DT model for a specific runtime context (\cite{DBLP}). In this work, we estimate the non-static control parameters on the real system by sampling the xDT, in presence of noise and uncertain model parameters. We then gradually move towards parameters that improve performance, by combining stochastic gradient approximation in simulation with data-driven approaches in the form of Unscented Kalman Filter. The adaptation method requires low computational effort and low data storage on memory, making it interesting for embedded hardware deployment real-time execution. It allows the control and planning strategies to adapt to a changing environment, situations, and driving styles by automatically tuning and calibrating the parameters. The proposed framework can deal with both unseen and edge-case scenarios by learning adequate control parameters without carrying an overhead of historical data and explicit policies. By leveraging simulation-based optimization methods and data-driven approaches we reduce tuning efforts and time as we avoid trial and error tuning campaigns. This method can improve testing automation in the automotive industry for tuning and validation of a controller or planner for Advanced Driver Assistance Systems (ADAS).

The paper is organized as follows. Section II discusses some background on sim2real methods for autonomous systems, as well as two gradient-free methods used to optimize over the controller parameters in the real world. Section III presents the implementation and results of the proposed methodology in automatic tuning for an autonomous driving application. We extend the implementation with a combination of two automatic calibration methods in Section IV, and discuss their convergence.%
\section{Background}%
There is a deep belief that increasing the simulator or the digital twin's accuracy alone, will not decrease the gap between simulation and reality. \cite{DBLP:journals/pami/MuratoreG021} discuss an SOB in machine learning and specially in trained models with reinforcement learning. The maximized expected return is generally overestimated in simulation. This causes the trained algorithm to be less performant, less robust, and more prone to failure in the real world. In this section, we quickly introduce the concepts of domain randomization and adaptation, and we discuss the xDT. Furthermore, we give a short technical background about gradient-free methods that could be used for automatic tuning. Finally, we formulate the path following NMPC.

\subsection{Sim2Real in Autonomous Systems}
To prevent overfitting and over-optimizing the algorithm in simulation, a technique called Domain Randomization (DR) is introduced. It consists of adding perturbation to the simulator's physical parameters (e.g mass, inertia, length, friction coefficients), noise to the control actions and state estimations, and disturbances to the visual properties, in training. This method regularizes the algorithm and robustifies it against real-world uncertainties. The agent trains to maximize the expected return over a distribution of uncertain parameters, delays, and noise levels. Another method of tackling sim to real problems is Domain Adaptation (DA). This transfer learning method  tackles the ability to train an algorithm in one source domain and deploy it on another, possibly different, target domain. This method learns to match the distributions in the target and source domain to reach a domain-invariant feature representation, which is used as an input to the trained agent.

To our knowledge, there is no single framework that combines the benefit of the three methods (DR, DA, and high-fidelity simulations) into one, allowing a simple transfer from simulation to reality. We introduce a learning method to learn the distribution in the domain randomization and update it on the fly. This could could be beneficial when parameters are not well defined in advance (system identification) and in the framework of learning the hyperparameters of a learning algorithm (meta-learning in \cite{DBLP:journals/corr/abs-2201-03916}). The goal is not to find parameters solving a single task well but rather generalizing to several tasks, with automatically adapted parameters.%
\subsection{Executable Digital Twin}%
Simulations are a main pillar in an X-in-the-loop (XiL: Model-in-the-loop, Software-in-the-loop, Hardware-in-the-loop, Vehicle-hardware-in-the-loop) framework. MiL is a major requirement in the automotive industry for the development and validation processes of control and planning algorithms as they allow to minimize the risk and effort in real-life testing. Reliable simulations can contribute to reducing the real mileage needed for algorithm validation (\cite{sontra2017}). 
There is also a trade-off between analytical models, which are often simple and differentiable, and black-box models of high-fidelity simulators. The DT integrates data and other knowledge of a physical asset and bridges the physical and the virtual world to assess performance, predict the behavior, and optimize the service (\cite{DBLP}). We use Simcenter Amesim to design the high-fidelity DT. Sim2real addresses the transfer of an algorithm from simulation with such DT to reality, without losing its performance, but more importantly, while keeping the real system stable and safe.
The xDT is an executable representation of a complex non-linear dynamical system or asset, in our case a road vehicle (\cite{DBLP}). It can be embedded on hardware, and run in a real-time environment. We make use of xDT to evaluate online the effect of control action and exploit learning-based controllers while easily varying model and environment parameters in an efficient, quick and safe way. According to \cite{DBLP} ‘‘[xDT] can be instantiated on the edge, on
premise, or in the cloud and used autonomously by a non-expert or a machine through [\dots] APIs’’. The xDT allows us to test configurations that are rather complex to recreate in real-life, to validate in XiL process \textcolor{black}{by exploiting dynamics and prediction that are often not captured well with simplistic analytical ODE car models}. Moreover, the xDT allows us to update the prior knowledge about our system online.%
\subsection{Gradient-free randomization methods}
\begin{figure}%
	\begin{center}
		\includegraphics[width=8.4cm]{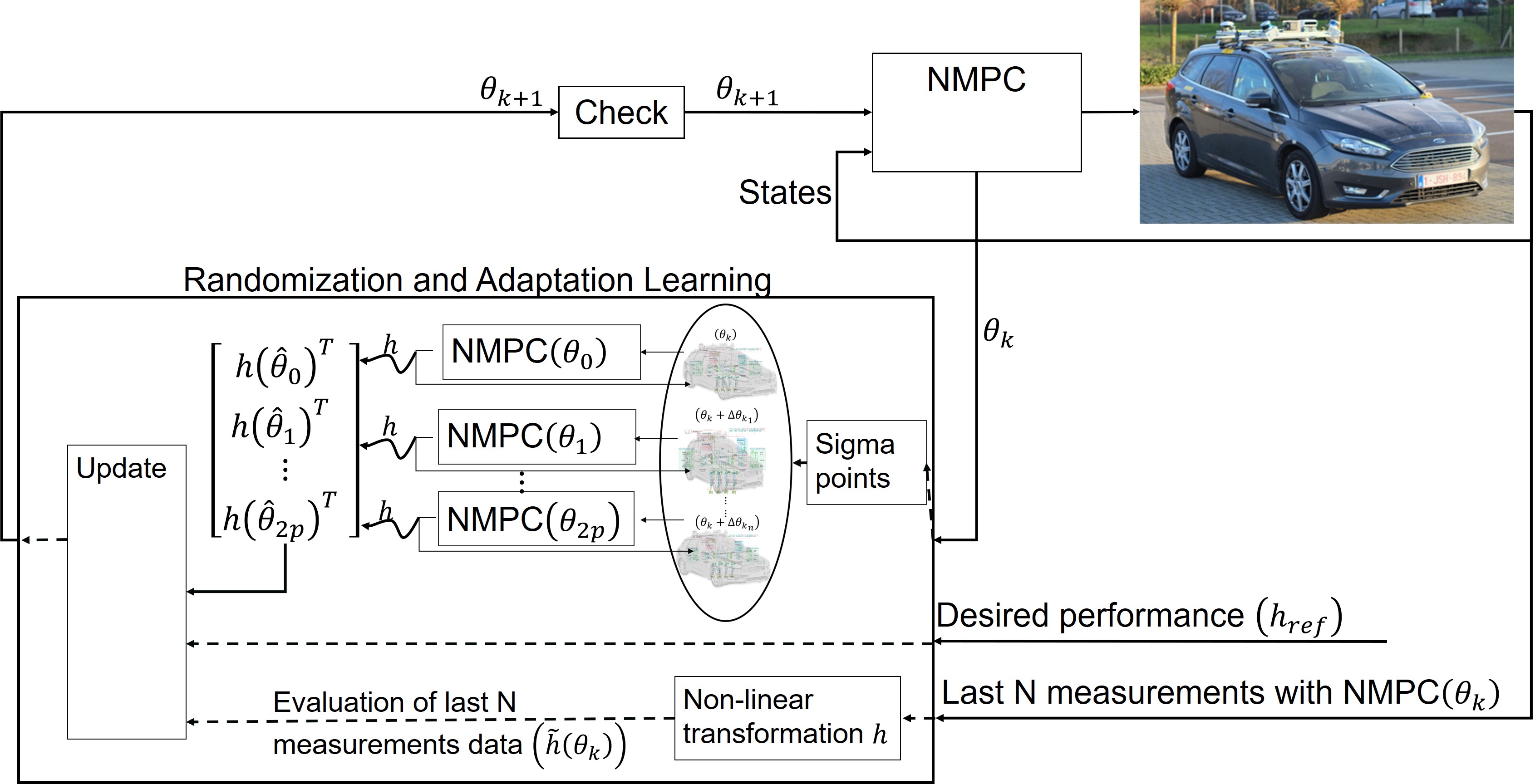}    
		\caption{Control parameter adaptation on an xDT: the main NMPC parameters are updated by sampling the xDT with NMPC($\theta_j$), evaluating the predicted performance $h_j$, and including the real measured performance $\tilde{h}$  of the last horizon N } 
		\label{fig:scheme_adaptive_controller}
	\end{center}
\end{figure}

We present two gradient-free methods that can be used to approximate the optimal set of control parameters on the real car. The adaptation schematic is shown in Figure~\ref{fig:scheme_adaptive_controller}. The NMPC parameters $\theta$ are adapted online to minimize the performance error $h_{ref} -  \tilde{h}(\theta_k)$, by sampling the xDT. We add randomization to $\theta_k$ and the model parameters.   %
\subsubsection{\textbf{Simultaneous Perturbation Stochastic Approximation (SPSA)}}%
\cite{Spall1998ANOO} introduces a stochastic approximation algorithm for stochastic optimization in multivariate systems. This method approximates the gradient of a loss function with respect to a system parameter with only two measurements, real or from simulation. Regardless of the system dimensionality, the gradient in the objective function can be approximated by perturbing all the parameters to be optimized over at once. For systems where the analytical relationship between an objective function and the parameters is unknown or difficult to develop, this method could be of significant benefit. This recursive optimization method significantly cuts down the number of iterations needed to estimate a gradient. SPSA seeks to minimize a loss function $L(\theta)$ where $\theta$ is a $p$-dimensional vector. We approximate the gradient $g(\theta) = \nabla L(\theta)$ as direct measurements of it are assumed non viable. The simultaneous perturbation step has all $p$-elements of $\theta_k$ perturbed at once. At an instance $k$, we run two perturbed simulations out of which two measurements of $L$ are sufficient to calculate the gradients with respect to every $j=1,\dots,p$ parameter such that:
\begin{equation}
	\hat{g}_{kj}(\hat{\theta}_k) = \frac{\partial L}{\partial\theta_{kj}} =\frac{L(\theta_k + c_k\Delta_k) - L(\theta_k - c_k\Delta_k)}{2c_k\Delta_{kj}}  
\end{equation} 
Every parameter is independently perturbed with a magnitude of $c_k\Delta_{kj}$ where $c_k$ is the differential step size hyper-parameter of the SPSA algorithm. \cite{Spall1998ANOO} suggests a random perturbation vector $\Delta_k = [\Delta_{k1},\dots,\Delta_{kp}]^T$, following a Bernoulli distribution symmetric about zero, with mutually independent elements. As opposed to other gradient-based methods requiring $p$ simulations, SPSA requires only 2, leading to considerable savings when $p$ is large. The gradient is approximated as:
\begin{equation}
	\hat{g}_{k}(\hat{\theta}_k) = \frac{L(\theta_k + c_k\Delta_k) - L(\theta_k - c_k\Delta_k)}{2c_k}
	\begin{pmatrix}
		\Delta_{k1}^{-1}\\
		\vdots\\
		\Delta_{kp}^{-1}
	\end{pmatrix}
\end{equation}
The general form in a recursive SPSA algorithm to solve for $\partial L/\partial\theta_k = 0$, with a step size $a_k$, is:%
\begin{equation}\label{eq:update_param_SPSA}%
	\hat{\theta}_{k+1} =  \hat{\theta}_k - a_k\hat{g}_k(\hat{\theta}_k)
\end{equation}%
\subsubsection{\textbf{Unscented Kalman Filter}}%
\begin{figure}
	\begin{center}
		\includegraphics[width=8.4cm]{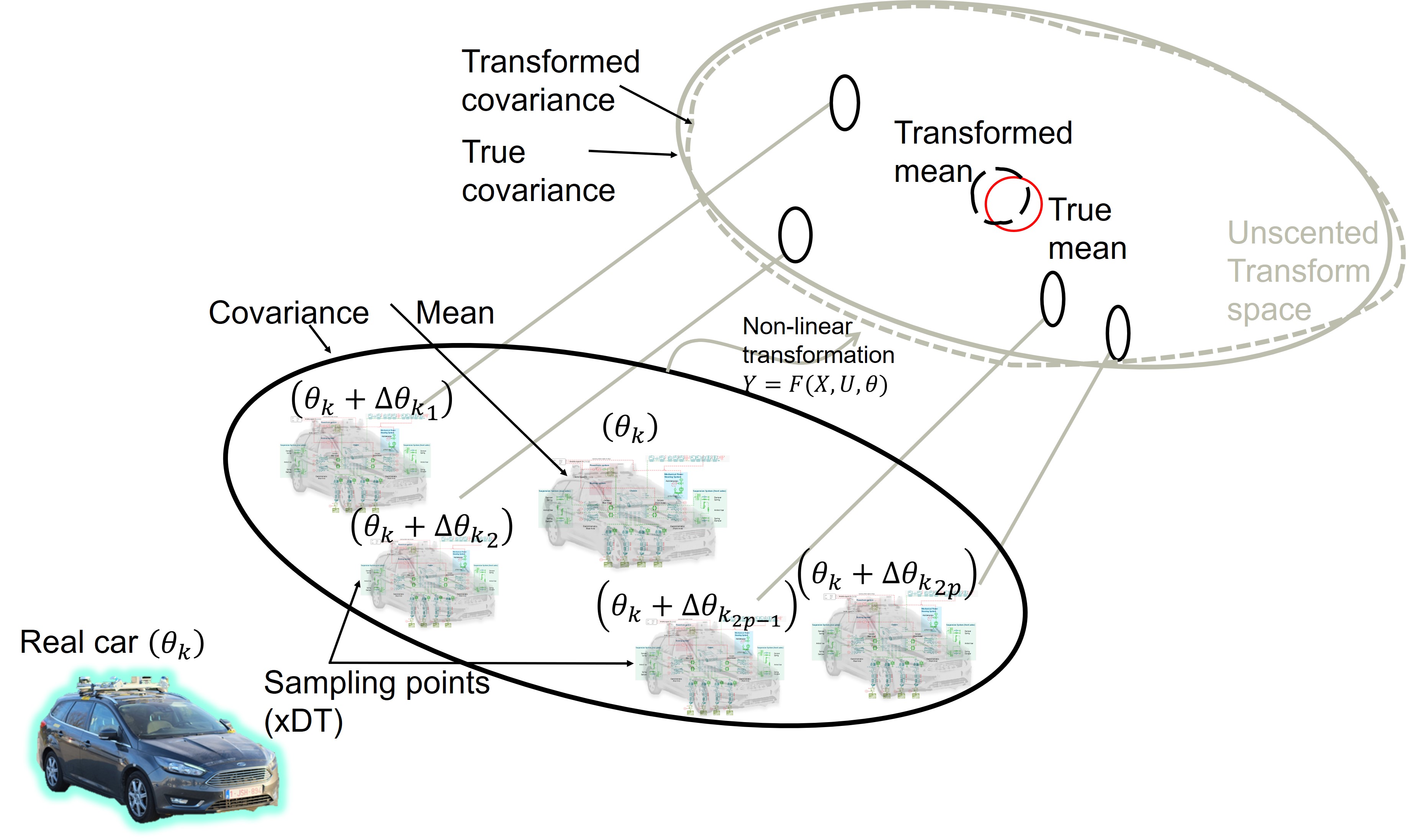}    
		\caption{Sampling points in an UKF: The mean and covariance of the transformed parameter distribution are approximated by sampling the xDT with $2p$ points around the current parameter set $\theta_k$ and measuring the respective outputs} 
		\label{fig:states_output_adaptiveukf}
	\end{center}
\end{figure}
In the work of \cite{menner2021automated}, an Unscented Kalman Filter method is employed to estimate the control parameters in PID, state-feedback, neural networks, and optimal controller,s and validate it with a vehicle simulator. An Unscented Kalman Filter propagates the parameters through non-linear dynamics and updates the estimated parameters without relying on gradient measurements or back-propagation. The method samples a set of points around the mean called sigma points, which are used to predict the output of the model. However, their sampling method predicts the system output using a simple bicycle model simulator. In our approach, we propose the closed-loop propagation of the perturbed parameters through a high-fidelity xDT. The predicted performance is closer to the real-world data, contributing to a quicker and safer convergence of the approach given a small sim2real reality gap. In addition, to add domain randomization, the model parameters in the xDT are perturbed and follow a normal distribution.  Contrary to the extended Kalman filter (EKF), the UKF does not linearly approximate the dynamics and output functions around the mean of the Gaussian to predict the values. This makes it attractive in a framework containing high-fidelity black-box models and non-linear evaluation metrics. For this, we first define the governing dynamics:
let $\mathcal{F}$ be a possibly non-linear transformation defining the system dynamics and stacking the states from time $k$ to $k+N$ where $N$ is a chosen horizon. $\mathcal{H}$ is a non-linear, possibly non-analytical and non-differentiable output transformation, serving as an evaluation metric map. $V$ and $N$ are the process and output noise.%
\begin{subequations}\label{eq:stochastic_dynamics_equation}%
	\begin{align}
		X_{k|k+N+1} &= \mathcal{F}(X_k,U_{k|k+N}, V_{k|k+N}, \theta_k)\\
		Y_{k|k+N}  &= \mathcal{H}(X_{k|k+N}, U_{k|k+N}, \theta_k) + N(k|k+N)
	\end{align}
\end{subequations}
Let $Y_{ref,k}$ be the stacking of reference evaluation metric from time $k-N$ to $k$. In our approach, $f(x, u, v, \theta)$ is the output of a black-box, xDT of the real vehicle. Given a $p$-sized parameter vector $\theta$ following a normal Gaussian distribution of the form $\theta \sim \mathcal{N}(\theta_k,\,P_{k|k})$  at an instance $k$, we obtain a set sigma or sampling points $\Theta$ around the mean $\Theta =[ \theta_k  \vert \theta_k + c_kA^{j}  \vert \theta_k - c_kA^{j}]\in \mathbb{R}^{p\times 2p+1}$:%
\begin{subequations}\label{eq:ukf_sampling_pts}%
\begin{align}
	\Theta^0 &= \theta_k,\\
	\Theta^j &= \theta_k + c_kA^j, 	   \quad \;\;\;\; j=1,\dots,p\\
	\Theta^j &= \theta_k - c_kA^{j-p}, \quad j=p+1,\dots,2p
\end{align}
\end{subequations}
where $c_k = \sqrt{p + \lambda}$, and $A^j$ is the $j$th column of the matrix $A = \sqrt{P_{k|k}}$. Matrix $A$ can be computed by performing a Cholesky decomposition of the prior covariance matrix such that $P_{k|k} = AA^T$.
The hyper-parameter $\lambda$ dictates the spread of the sampling (sigma) points around the mean. In total, $2p+1$ time evolutions are performed for one update step.
The Unscented Transformation is based on the weighted mean of all the sigma points and their covariance. We form the weighting vector $\mathcal{W} = [w_a^0,\, w_a^1,\, \dots,\, w_a^{2p}] \in \mathbb{R}^{1\times 2p+1}$  such that:%
\begin{align}\label{eq:ukf_weighting_pts}%
	w_a^0 = \frac{\lambda}{(p + \lambda)}, \,
	w_a^j = \frac{1}{2(p+\lambda)} \quad \textrm{for } j=1,\dots,2p,
\end{align}
where $w_a^j$ is the weight associated
with the $j$th point, and $w_a^0$ is the weight associated with the first sigma point, which is the mean $\theta_k$. A good heuristic for choosing $\lambda$ according to \cite{847726}, would be $p + \lambda = 3$.
\begin{algorithm}[t]
	\caption{Automatic parameter estimation algorithm}\label{alg:ukf_parameter_estimation}
	\begin{algorithmic}
		\Require $\Theta, w_a,  C_{\theta}, C_n$
		\Function{Unscented Transform}{} 
			\State $\bar{\theta} = \sum_{j=0}^{2p}w_a^j\Theta^j$
			\State 	$P_{k+1|k} = C_\theta + \sum_{j=0}^{2p}w_a^j(\Theta^j - \bar{\theta})(\Theta^j - \bar{\theta})^T$
			\State $\mathcal{Y}^j = h(\theta^j, x_{k-N})$\Comment{Propagate}
			\State $\bar{y} = \sum_{j=0}^{2p}w_a^j\mathcal{Y}^j$
 		 \EndFunction

		\Function{Measurement update step}{}
			\State $P_{\theta y} = \sum_{j=0}^{2p}w_a^j(\Theta^j - \bar{\theta})(\mathcal{Y}^j - \bar{y})^T$
			\State $P_y = C_n + \sum_{j=0}^{2p}w_a^j(\mathcal{Y}^j - \bar{y})(\mathcal{Y}^j - \bar{y})^T$
		 \EndFunction
	\State $K_k = P_{\theta y}P_y^{-1}$ \Comment{Kalman gain}
	\State $P_{k+1|k+1} = P_{k+1|k} - K_kP_yK_k^T$ \Comment{Posterior covariance}
	\end{algorithmic}
\end{algorithm}
The step $\mathcal{Y}^j = h(\theta^j, x_{k-N})$ in algorithm~\ref{alg:ukf_parameter_estimation} propagates the sigma points through the non-linear transformations (dynamics and evaluation) starting with the initial condition state $x_{k-N}$ until time $k$. This method seeks to find the set of control parameters that would have improved the performance from time $k-N$ to $k$ where $N$ is the length of the data from the real system.
The control parameters are then updated according to algorithm \ref{alg:ukf_parameter_estimation} and the law:%
\begin{equation}
	\theta_{k+1} = \theta_k + K_k(\mathcal{Y}_{ref,k} - \tilde{h}(\theta_k)),
\end{equation}%
where $\tilde{h}(\theta_k)$ is the vector-valued evaluation metric from the real system (\cite{menner2021automated, 882463}). $C_n$ and $C_\theta$ are output and process noise covariance matrices.
\subsection{NMPC: Path following formulation}
\begin{figure}%
	\centering
	\subfloat[\centering Single track model]{{\includegraphics[height=2cm]{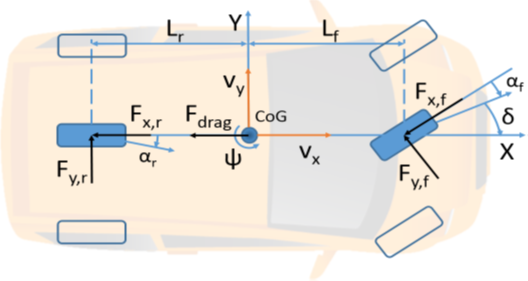} }}%
	\subfloat[\centering Curvilinear frame transformation]{{\includegraphics[ height=2.5cm]{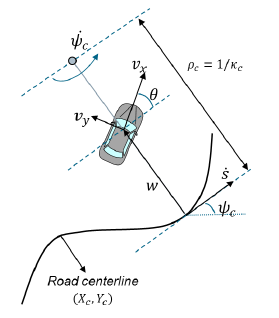} }}%
	\caption{Single track curvilinear model for path following}%
	\label{fig:single_track_curvilinear}%
\end{figure}

Car dynamics used in the NMPC are represented with a real-time feasible model such as the 6 DoF single-track model described in \cite{DBLP:journals/corr/abs-2110-03349}, satisfactory in path following and lane keeping scenarios.%
{\begin{align}
		\label{eq:BicycleModelStates}
		\Dot{v}_x &= (F_{xf} cos\delta + F_{xr} -F_{yf} sin\delta - F_{res} + M\Dot{\psi} v_y)/M, \nonumber\\
		\Dot{v}_y &= (F_{xf} sin\delta + F_{yr} + F_{yf} cos\delta - M\Dot{\psi} v_x)/M, \\\nonumber
		\Dot{r} &= (L_f (F_{yf} cos\delta + F_{xf} sin\delta) - L_r F_{yr})/I_z, \\\nonumber
		\Dot{s} &= (v_x \cos\theta -v_y sin\theta)/(1-\kappa_c w), \\\nonumber
		\Dot{w} &= v_x \sin\theta + v_y \cos\theta, \\\nonumber
		\Dot{\theta} &= \Dot{\psi} - \Dot{\psi_c} = \Dot{\psi} - \kappa_c \Dot{s}.\\\nonumber
\end{align}}%
In the single-track vehicle model of \eqref{eq:BicycleModelStates} and Figure~\ref{fig:single_track_curvilinear}, the first three equations dictate the dynamics in the car body frame with $x$ pointing forward and last three dictate the kinematics in the curvilinear frame. We add the state $s$ to track the evolution along the center-line. Inputs to the model are the body frame steering angle $\delta$ and the front/rear axles longitudinal forces $F_{xf}$ and $F_{xr}$. At moderate speeds, a linear tire model estimates the lateral forces $F_{yf}$ and $F_{yr}$ as proportional to the slip angles $\alpha_f, \alpha_r$, assuming small slip angles using constant cornering stiffness.
For control purposes, it is beneficial to represent the car position with respect to the track and path, in the curvilinear reference frame, local and attached to the car body frame as in Figure~\ref{fig:single_track_curvilinear}(b). Each point on a path is defined by a 4-tuple $X_c, Y_c, \psi_c, \kappa_c$ to represent the position, heading, and a curvature in the Cartesian frame. The transformation to a curvilinear frame facilitates the optimal control problem (OCP) formulation as it is only function of the road curvature. Finally, track limits can be added as varying box constraints as $w_l$ and $w_r$, the left and right distances from the center-line. $w$ and $\theta$ are respectively the distance and heading deviation from the center-line, with $w>0$ to the left of the center-line and $\theta>0$ a counter-clockwise rotation with respect to the tangent to the center-line $\psi_c$.%
\begin{align}%
		\label{Eq:CartToCurv}
		\begin{split}
			w &= (Y-Y_c)\cos(\psi_c) - (X-X_c)\sin(\psi_c),\\
			\theta &= \psi - \psi_c. 
		\end{split}
\end{align}%
%
To summarize, the single-track dynamics between the state vector $x$ and input $u$, in the Curvilinear frame is:%
\begin{equation}%
	\label{eq:CartesianStatesControl}
	\begin{split}
		\Dot{x} = f(x,u),\;
		x = [v_x, v_y, r, s, w, \theta ],\;  u = [\delta, t_r].
	\end{split}
\end{equation}%
NMPC controls the car by computing the normalized throttle $t_r$, and the steering angle $\delta$. We use a receding horizon scheme of $N_H$ steps, and we augment the dynamics with the input rates variables $\dot{u}$ for smoother driving. Moreover, we obtain the nonlinear difference equations $f_d$ by applying a 4 step Runge-Kutta $4^{th}$ order method to the dynamics $\Dot{x} = f(x,u)$ in~\eqref{eq:CartesianStatesControl}. Finally, the NLP optimizes over the discrete-time OCP for path following as in~\eqref{eq:NMPC_Formulation}.%
{\small\begin{align}\label{eq:NMPC_Formulation}%
				\min_{x(0),\dots, x(N),u(0),\dots, u(N-1)} \sum_{k=0}^{N-1}  l_k(x_{k} -&x_{ref,k},u_{k}) + V_{N}(x_{N})\\\nonumber
				\textrm{subject to: }  x_{0}                            = x(0) \quad&\textrm{(Initial condition)}\\\nonumber
		 x(k+1) = f_d([x(k),u(k)], \dot{u}(k))                    \quad &\textrm{(Dynamic equations)} \\ \nonumber
		x_{min} \leq x(k) \leq x_{max}                            \quad &\textrm{(State constraints)}\\\nonumber
		u_{min} \leq u(k) \leq u_{max}                            \quad &\textrm{(Input constraints)}\\\nonumber
  \dot{u}_{min} \leq \dot{u(k)} \leq \dot{u}_{max}                   \quad &\textrm{(Input rate constraints)}
\end{align}}%
$V_N(x_N)$ is a terminal cost and the stage cost $l_k(x_{k},u_{k})$ is:%
{\small\begin{align}\label{eq:stage_cost_nmpc}%
		l_k(x_k, u_k) = x(k) ^ T Q x(k)
		 + u(k) ^ T R u(k) + \dot{u}(k) ^ T S\dot{u}(k)^T, 
\end{align}}%
with $Q \in\mathbb{R}^{6\times 6} \succeq 0, R \in\mathbb{R}^{2\times 2} \succ 0, S\in\mathbb{R}^{2\times 2} \succ 0$. The path following problem reduces to a regulation about a zero reference for all states except the velocity.

\section{Implementation for Autonomous Driving Applications}
In order to demonstrate the benefit of an automatic adaptation sampling an xDT, we implement our methodology on an autonomous driving application. While our approach is not restricted to tuning controllers, or to a specific type of controllers, we demonstrate it on a path following NMPC. This section formulates a performance-improver automatic calibrator.
We build on the XiL verification process described in the work of \cite{DBLP:journals/corr/abs-2110-03349} for the deployment of a real-time NMPC on embedded platforms. The NMPC is solved to full convergence using a Sequential Quadratic Programming (SQP) method with an active-set QP solver. We optimize over the resulting NLP using a multiple-shooting framework, with a sampling time $T_s = 40ms$ and an NMPC horizon $N_H = 30$.
\subsection{Problem Formulation}
The proposed automatic adaptation method can calibrate controllers of an unknown dynamical system online, rendering it attractive for embedded deployment in a real-world system as a sim2real method.
For our implementation, the controller learns to track a parameterized reference path and velocity profile, in 4 successive lane changes. The function $\mathcal{H}$ in~\eqref{eq:stochastic_dynamics_equation} converts the measured output over the past time window of size $N$, to a vector-valued performance metric as shown in the framework of Figures \ref{fig:scheme_adaptive_controller} and \ref{fig:states_output_adaptiveukf}. As the single-track dynamics are formulated in a curvilinear frame with respect to the path, deviation from the path center-line is the state $w$. We then form the performance vector $h(\theta_k)$ such that:  
\begin{equation}
	h(\theta_k) = 
	\begin{bmatrix}
		10(\mathcal{V}_x - \mathcal{V}_{ref})\\
		10(w_{k-N|k})\\
		\mathcal{J}^*_{NMPC}
	\end{bmatrix},
	\mathcal{Y}_{ref,k} = 
\begin{bmatrix}
	0\\
	0\\
	0
\end{bmatrix},
\end{equation}
where $\mathcal{V}_x - \mathcal{V}_{ref} = v_{x,k-N|k} - v_{ref,k-N|k}$ is a vector stacking of velocity tracking errors over the past time window. Similarly for the deviation from the path $w_{k-N|k}$. In addition, we include the stacking of the NMPC's optimal cost $\mathcal{J}_{NMPC}^*$ such that the algorithm minimizes it to keep the energy bounded.  The factor of $10$ in the tracking errors is added for normalization.
The objective performance in this implementation leads to a tradeoff between reducing $Q,R,S$ such that $\mathcal{J}_{NMPC}^* \rightarrow 0$, and increasing $Q,R,S$ to minimize the velocity and path tracking errors. Another way of implementing this, is to set an activation function:
\begin{equation}
	r(J^*) = \max(J^*, \underbar{J}) - \underbar{J}
\end{equation}
where $\underbar{J}$ is a threshold for an allowable optimal cost that keeps the system stable. For this particular implementation, we are interested in tuning the controller parameters, specifically the $Q$, $R$, and $S$ matrices in~\eqref{eq:stage_cost_nmpc}. This method can be also used to optimize offline over the horizon length $N_H$ of the NMPC resulting in the best performance and computation time.
Increasing the weights on the OCP variables could render the behavior slightly more aggressive and dynamic, and potentially increase the number of SQP and QP iterations needed to solve for the primal and dual estimates. Therefore, we can augment the vector-valued objective metric by the NMPC computation time, to reach a trade-off between tracking performance and real-time feasibility. Adding a performance metric such as the computation time helps demonstrate the benefit of this method in dealing with black-box, non-analytical, and non-differentiable by design, choice of measurements.
We choose a starting hyperparameter $\lambda = 1$, $w_0  = 1/3$ and a horizon length of 3 seconds to update the controller.%
\subsubsection{Regularization heuristic}%
As the sampling points are calculated in the Unscented Transform step, it could occur that the algorithm tries to run a simulation with indefinite $Q$, $R$, $S$ matrices resulting from negative parameters in $\theta$. While \cite{menner2021automated} argues that the method learns to stay away from indefinite matrices, we add a regularization process that guarantees the positive definiteness of the parameters. As we sample with an xDT, this allows avoiding infeasible and unstable roll-outs. The regularization finds the smallest step size $c$ that causes the sampling point to activate the high or low boundary constraints $\bar{\theta}$ and $\underbar{$\theta$}$. This regularization method does not clip the parameters individually to their limits, but rather finds a coefficient that conserves the shape of Gaussian distribution around the mean. The coefficient is calculated such that: %
$	c_k = \min(c_0, c|
\underbar{$\theta$} \leq \theta_k + cA^j \leq \bar{\theta} \,\, \& \,\,
\underbar{$\theta$} \leq \theta_k - cA^j \leq \bar{\theta}),
$
where $c_0 = \sqrt{p + \lambda}$.
We also update the process noise covariance matrix using real-world data.
$C_n$ is updated using the stacking of the measured slack variables $\tilde{N} = \mathcal{Y}_{ref,k} - h(\theta_k)$ from real measurements.



\subsection{Example of successive double lane changes}
\begin{figure}
	\begin{center}
		\includegraphics[width=8.0cm]{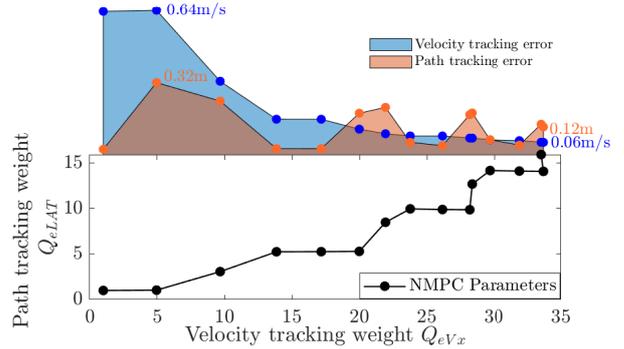}    
		\vspace{-8pt}
		\caption{Automatic tuning without noise: Control parameter evolution and Infinity norm of tracking error  } 
		\label{fig:parameter_update_ukf}
	\end{center}
\end{figure}
\begin{figure}
	\begin{center}
		\input{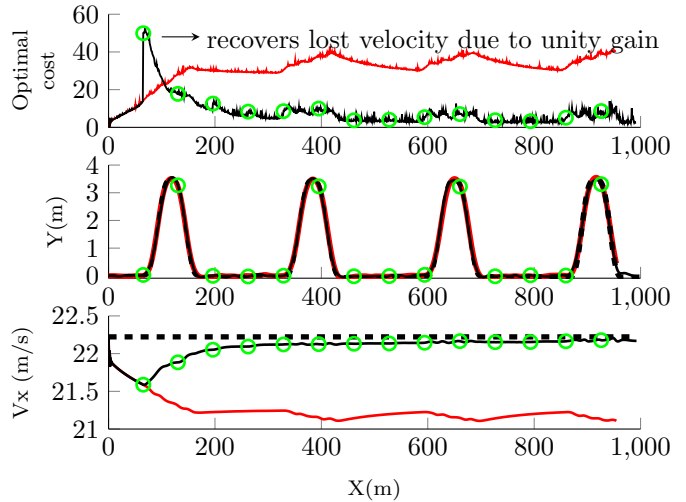}    
		\caption{Convergence of NMPC cost function (with automatic tuning: solid-black, without automatic tuning: red, reference: dashed-black, tuning instances: green)} 
		\label{fig:cost_states_evo}
	\end{center}
\end{figure}

We validate our methodology with 4 successive ISO 3888-1 standard double lane change scenarios at 80kph. We show how such a formulation could facilitate the task of controller adaptation. We start with a unit and diagonal $Q$, $R$, and $S$ matrices, with no prior knowledge.  For sake of simplicity, we visualize two elements of the $Q$ matrix, namely $Q_{eLAT}$ and $Q_{eVx}$ which are the weights on the lateral deviation error and velocity tracking error respectively, as seen in Figure~\ref{fig:parameter_update_ukf} along with the infinity norm of the tracking errors. Every dot represents newly updated parameter set, evolving with time from left to right. At the end of the scenario, with only one run, the algorithm adapts the control parameters such that the velocity tracking error drops by almost 91\% and the path tracking error by 63\%. From the second update step, the algorithm detects the model mismatch in the longitudinal dynamics between the single-track model and the xDT and increases the weight on the velocity to 10.
This plot allows us to visualize the sensitivity of the tracking error norm to the change in parameters. In those sections with little to no steering (Figure~\ref{fig:cost_states_evo}), the algorithm does not learn about the lateral deviation. The automatic tuning updates $Q_{eLAT}$ mainly during the lane change, as seen in the peaks of the path tracking error in Figure~\ref{fig:parameter_update_ukf}. Every peak corresponds to a double lane change, and it is clear that the error peak norm decreases as the scenario evolves.

Figure~\ref{fig:cost_states_evo} shows the closed-loop result of the scenario with and without automatic tuning using an xDT. Green dots correspond to an update instance (every 3 seconds). In the first three seconds, as the gains are set to unity, the velocity drops considerably, to which the algorithm reacts by increasing the weight on the velocity tracking error. This results in an increase in throttle, all while still satisfying the constraints of the NMPC. The switch leads to a peak in the optimal cost but it is followed by a quick decrease. This shows the potential of this method, as it improves the tracking performance and keeps the NMPC optimal cost bounded.

\section{Results and discussion}
As we deal with an adaptive controller, it is important to add safety checks to ensure that the system is not diverging, and to verify the applicability of the updated control parameter set. In this section, we discuss an energy-based study to validate the controller. Moreover, we compare the UKF-data-driven approach to a simulation-based-optimization using SPSA. Finally, we combine both methods and show potential results.

\subsection{Convergence analysis}
For applications with highly non-linear dynamics and in presence of noise, theoretical Lyapunov stability is hard to prove. However, the xDT allows us to quickly verify the new controller and calculate its associated benefit in energy, if we can find a function $V(x) = x^TPx$ where $P \succ 0$ such that the $V(x(k+1)) \le V(x(k))$. In a less strict sense, given the noise sources, we aim to find a non-monotonically decreasing energy function contained within a ball of radius $R$.
The first controller check is such that $V(P_{k+1}, x_{k-N|k}) \leq V(P_{k}, x_{k-N|k})$ where $P_k = diag(\theta_k) \succ 0$, is the diagonal matrix constructed from the entries of the tuned controller parameters: the estimated $\theta_{k+1}$ would have reduced the energy function if it were applied instead of $\theta_{k}$ starting $x(k-N)$. The second validation is through applying the updated controller parameter to the real vehicle and verifying that the resulting energy function in $[k,\dots,k+N]$ is non-increasing.



SPSA perturbs the parameter (sampling points) with $\Delta_{k,i}$ following a Bernoulli distribution with $p = 0.5$, with all p-parameters being mutually independent. We set $a_k=0.05, c_k=0.1$ as we try to estimate non-static parameters. The parameters are perturbed and propagated through the xDT, similar to the proposed tuning method in the previous section. We add an additive output noise with a signal-to-noise ratio of 2dB. To enhance the gradient approximation in presence of noise, we run $2p$ sets of perturbed parameter simulations, symmetric with respect to 0 such that $\Delta_{k,i} =  \pm 1$  and $\Delta_{k,i} =  \pm 2$. We then average the approximated gradients from simulation and update the control parameter according to~\eqref{eq:update_param_SPSA}. However, as this method performs a simulation-based optimization, it does not take into consideration real world data and can perform at best to compensate for noise and mismatch between the NMPC's single-track model and the xDT. We add it to our approach as a comparison.
To leverage simulation and real-world data, we combine both methods. The UKF technique propagates the dynamic with $2p+1$ set of parameters, sampled according to the prior knowledge on distribution $P_{k|k}$ and noise level. We make use of the measured performance metric $\mathcal{Y}$ to approximate the gradient around the current $\theta_k$ using SPSA. We then average the steps $\Delta\theta_k$ from both methods. The SPSA method optimizes over a scalar loss function $L$.
We set $L(\theta_k) = \|h(\theta_k)_{1:N}\|^2_2 + \|h(\theta_k)_{N:2N}\|^2_2 $ where $\|h(\theta_k)_{1:N}\|_2$ is the $2$-norm of the $h$ vector containing the velocity tracking error and $\|h(\theta_k)_{N:2N}\|_2$ contains the path tracking errors.
The SPSA method is quicker to react to the path tracking error compared to the UKF, and is less sensitive to the velocity tracking error as seen in Figure~\ref{fig:infinitynorm_errors_v2}. It also results in a smaller tracking error. However, combining both methods results in a quicker reaction, from the first iteration, and simultaneously to both tracking errors showing the benefit of updating the estimations with real data, all while employing the DT to optimize over the parameters. Overall, velocity tracking error drops from $1m/s$ to around $0.1m/s$ for the SPSA, UKF and UKF+SPSA. Infinity norm on the path tracking drops from 0.26m to 0.12m with UKF+SPSA. Both methods have similar trends and effects on the NMPC optimal cost. Another important aspect to consider is the noise effect as we compare the UKF in Figures~\ref{fig:parameter_update_ukf} and~\ref{fig:Q_evo_ukf_spsa_ukfwspsa_v2}. As the algorithm has prior and posterior knowledge about the noise level, it adapts less aggressively to tracking error measurements which are mainly due to noise. $Q_{eLAT}$ is 15 without noise and 5 with noise at the end of the scenario. 
The control parameters adaptation is shown in Figure~\ref{fig:Q_evo_ukf_spsa_ukfwspsa_v2}.
\begin{figure} 
	\begin{center}
		\input{Figures/8_Q_evo_ukf_spsa_ukfwspsa_v4_tex}    
		\caption{Automatic tuning: control parameters evolution (UKF: black, SPSA: grey, UKF+SPSA: blue)} 
		\label{fig:Q_evo_ukf_spsa_ukfwspsa_v2}
	\end{center}
\end{figure}
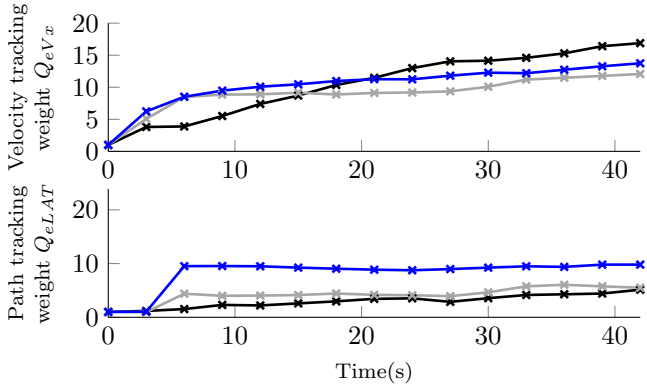
\begin{figure}
	\begin{center}
		\input{Figures/9_infinitynorm_errors_v4_tex}    
		\caption{Automatic tuning: infinity-norm tracking error evolution (UKF: black, SPSA: grey, UKF+SPSA: blue, No tuning: red)} 
		\label{fig:infinitynorm_errors_v2}
	\end{center}
\end{figure}
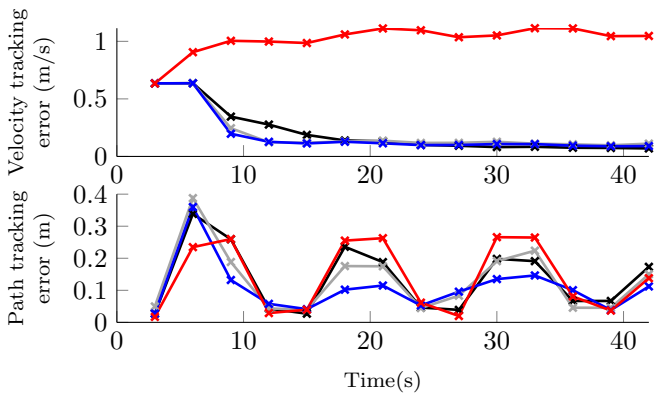



\section{Conclusion}
This paper presents a learning-based adaptation framework for transferring control and planning strategies from simulation to the real setup without manual tuning. The adaptation occurs online and on-the-go as the optimal parameters are estimated to minimize an objective performance error. The proposed algorithm automatically tunes the parameters, to unseen noise levels, edge-case situations and environment, by closing the loop on the prediction using an executable digital twin with real-world data. Unscented Kalman Filter is used to sample a set of perturbed parameters and propagate them through complex non-linear predictive systems. SPSA minimizes the performance error by approximating the performance gradient with respect to the parameters using only 2 simulations regardless of the dimension of the parameters. We validate the proposed algorithm on 4 successive lane changes and show that it estimates the parameters in only one online run, improving the performance up to 91\%.
\begin{ack}
This project has received funding from the European Union’s Horizon 2020 research and innovation programme under the Marie Skłodowska-Curie grant agreement ELO-X No 953348
\end{ack}

\bibliography{ifacconf} 
%

\end{document}

%% file: Figures/8_Q_evo_ukf_spsa_ukfwspsa_v4_tex.tex
%
%
\begin{tikzpicture}

\begin{axis}[%
	width=7.0cm,
	height=1.7cm,
	at={(0.0cm,2.2cm)},
	scale only axis,
	xmin=0,
	xmax=42,
	ymin=0,
	ymax=20,
	ylabel near ticks,
	ylabel={Velocity tracking \\ weight $Q_{eVx}$},
	ylabel style={font={\color{white!15!black}}, align=center},
	label style={font=\small},
	axis background/.style={fill=white},
	axis x line*=bottom,
	axis y line*=left
]
\addplot [color=black, line width=1.0pt, mark=x, mark options={solid, black}, forget plot]
  table[row sep=crcr]{%
0	1\\
3.00000000000009	3.80374\\
5.99999999999991	3.88535\\
8.99999999999989	5.5118\\
11.9999999999998	7.39808\\
14.9999999999998	8.7008\\
17.9999999999997	10.3549\\
20.9999999999996	11.4726\\
23.9999999999996	12.9864\\
26.9999999999995	14.0421\\
29.9999999999994	14.1401\\
33.0000000000008	14.5892\\
36.0000000000007	15.2932\\
39.0000000000007	16.4092\\
42.0000000000006	16.8805\\
};
\addplot [color=white!65!black, line width=1.0pt, mark=x, mark options={solid, white!65!black}, forget plot]
  table[row sep=crcr]{%
0	1\\
3.00000000000009	5.06883\\
5.99999999999991	8.47937\\
8.99999999999989	8.85748\\
11.9999999999998	8.89641\\
14.9999999999998	9.15618\\
17.9999999999997	8.86648\\
20.9999999999996	9.11787\\
23.9999999999996	9.1862\\
26.9999999999995	9.36565\\
29.9999999999994	10.0783\\
33.0000000000008	11.197\\
36.0000000000007	11.4818\\
39.0000000000007	11.7664\\
42.0000000000006	12.0723\\
};
\addplot [color=blue, line width=1.0pt, mark=x, mark options={solid, blue}, forget plot]
  table[row sep=crcr]{%
0	1\\
3.00000000000009	6.2354\\
5.99999999999991	8.51841\\
8.99999999999989	9.46793\\
11.9999999999998	10.0853\\
14.9999999999998	10.461\\
17.9999999999997	10.9715\\
20.9999999999996	11.2597\\
23.9999999999996	11.2385\\
26.9999999999995	11.8031\\
29.9999999999994	12.2749\\
33.0000000000008	12.1916\\
36.0000000000007	12.7315\\
39.0000000000007	13.2713\\
42.0000000000006	13.7386\\
};
\end{axis}

\begin{axis}[%
	width=7.0cm,
	height=1.7cm,
	at={(0.0cm,0.0cm)},
	scale only axis,
	xmin=0,
	xmax=42,
	xlabel near ticks,
	xlabel style={font=\color{white!15!black}},
	xlabel={Time(s)},
	ymin=0,
	ymax=23.8711,
	ylabel near ticks,
	ylabel style={font=\color{white!15!black}},
	ylabel={Path tracking \\ weight $Q_{eLAT}$},
	ylabel style={font={\color{white!15!black}}, align=center},
	label style={font=\small},
	axis background/.style={fill=white},
	axis x line*=bottom,
	axis y line*=left
]
\addplot [color=black, line width=1.0pt, mark=x, mark options={solid, black}, forget plot]
  table[row sep=crcr]{%
0	1\\
3.00000000000009	1.18342\\
5.99999999999991	1.53057\\
8.99999999999989	2.30174\\
11.9999999999998	2.20552\\
14.9999999999998	2.58631\\
17.9999999999997	2.96683\\
20.9999999999996	3.43408\\
23.9999999999996	3.52471\\
26.9999999999995	2.87805\\
29.9999999999994	3.55817\\
33.0000000000008	4.15373\\
36.0000000000007	4.28378\\
39.0000000000007	4.419\\
42.0000000000006	5.13908\\
};
\addplot [color=white!65!black, line width=1.0pt, mark=x, mark options={solid, white!65!black}, forget plot]
  table[row sep=crcr]{%
0	1\\
3.00000000000009	0.974505\\
5.99999999999991	4.38505\\
8.99999999999989	4.00695\\
11.9999999999998	4.04587\\
14.9999999999998	4.14066\\
17.9999999999997	4.43037\\
20.9999999999996	4.17898\\
23.9999999999996	4.11064\\
26.9999999999995	3.93119\\
29.9999999999994	4.64381\\
33.0000000000008	5.76253\\
36.0000000000007	6.04732\\
39.0000000000007	5.76269\\
42.0000000000006	5.50256\\
};
\addplot [color=blue, line width=1.0pt, mark=x, mark options={solid, blue}, forget plot]
  table[row sep=crcr]{%
0	1\\
3.00000000000009	1.03943\\
5.99999999999991	9.52262\\
8.99999999999989	9.54047\\
11.9999999999998	9.49423\\
14.9999999999998	9.24711\\
17.9999999999997	9.04125\\
20.9999999999996	8.865\\
23.9999999999996	8.7539\\
26.9999999999995	8.97598\\
29.9999999999994	9.24535\\
33.0000000000008	9.4922\\
36.0000000000007	9.3821\\
39.0000000000007	9.80267\\
42.0000000000006	9.80574\\
};
\end{axis}

\begin{axis}[%
	width=8.4cm,
	height=4cm,
	at={(0.0cm,0.0cm)},
	scale only axis,
	xmin=0,
	xmax=1,
	ymin=0,
	ymax=1,
	axis line style={draw=none},
	ticks=none,
	axis x line*=bottom,
	axis y line*=left
	]
\end{axis}
\end{tikzpicture}%

%% file: Figures/9_infinitynorm_errors_v4_tex.tex
%
%
\begin{tikzpicture}

\begin{axis}[%
	width=7.0cm,
	height=1.7cm,
	at={(0.0cm,2.2cm)},
	scale only axis,
	xmin=0,
	xmax=42,
	xlabel style={font=\color{white!15!black}},
	ymin=0,
	ymax=1.114,
	ylabel={Velocity tracking \\ error (m/s)},
	ylabel near ticks,
	ylabel style={font=\color{white!15!black}, align=center},
	label style={font=\small},
	axis background/.style={fill=white},
	axis x line*=bottom,
	axis y line*=left
]
\addplot [color=black, line width=1.0pt, mark=x, mark options={solid, black},  forget plot]
  table[row sep=crcr]{%
3.00000000000009	0.634399999999999\\
5.99999999999991	0.636100000000003\\
8.99999999999989	0.346\\
11.9999999999998	0.276700000000002\\
14.9999999999998	0.187100000000001\\
17.9999999999997	0.138999999999999\\
20.9999999999996	0.131500000000003\\
23.9999999999996	0.100100000000001\\
26.9999999999995	0.0926000000000009\\
29.9999999999994	0.0812000000000026\\
33.0000000000008	0.0837000000000003\\
36.0000000000007	0.0755000000000017\\
39.0000000000007	0.0737000000000023\\
42.0000000000006	0.0696000000000012\\
};
\addplot [color=white!65!black, line width=1.0pt, mark=x, mark options={solid, white!65!black}, forget plot]
  table[row sep=crcr]{%
3.00000000000009	0.634900000000002\\
5.99999999999991	0.635899999999999\\
8.99999999999989	0.2453\\
11.9999999999998	0.123200000000001\\
14.9999999999998	0.119\\
17.9999999999997	0.130400000000002\\
20.9999999999996	0.136700000000001\\
23.9999999999996	0.1174\\
26.9999999999995	0.118200000000002\\
29.9999999999994	0.127200000000002\\
33.0000000000008	0.110800000000001\\
36.0000000000007	0.1051\\
39.0000000000007	0.0970000000000013\\
42.0000000000006	0.111000000000001\\
};
\addplot [color=blue, line width=1.0pt, mark=x, mark options={solid, blue}, forget plot]
  table[row sep=crcr]{%
3.00000000000009	0.635100000000001\\
5.99999999999991	0.6356\\
8.99999999999989	0.197100000000002\\
11.9999999999998	0.127200000000002\\
14.9999999999998	0.113\\
17.9999999999997	0.127600000000001\\
20.9999999999996	0.114000000000001\\
23.9999999999996	0.0986000000000011\\
26.9999999999995	0.0996000000000024\\
29.9999999999994	0.107500000000002\\
33.0000000000008	0.107099999999999\\
36.0000000000007	0.0935000000000024\\
39.0000000000007	0.0884\\
42.0000000000006	0.0900999999999996\\
};
\addplot [color=red, line width=1.0pt, mark=x, mark options={solid, red}, forget plot]
  table[row sep=crcr]{%
3.00000000000009	0.6343\\
5.99999999999991	0.9057\\
8.99999999999989	1.0047\\
11.9999999999998	0.998200000000001\\
14.9999999999998	0.985400000000002\\
17.9999999999997	1.0599\\
20.9999999999996	1.112\\
23.9999999999996	1.0962\\
26.9999999999995	1.0356\\
29.9999999999994	1.0515\\
33.0000000000008	1.114\\
36.0000000000007	1.1124\\
39.0000000000007	1.0451\\
42.0000000000006	1.0469\\
};
\end{axis}

\begin{axis}[%
	width=7.0cm,
	height=1.7cm,
	at={(0.0cm,0.0cm)},
	scale only axis,
	xmin=0,
	xmax=42,
	xlabel style={font=\color{white!15!black}},
	xlabel={Time(s)},
	ymin=0,
	ymax=0.4,
	ylabel={Path tracking \\ error (m)},
	ylabel near ticks,
	ylabel style={font=\color{white!15!black}, align=center},
	label style={font=\small},
	axis background/.style={fill=white},
	axis x line*=bottom,
	axis y line*=left
]
\addplot [color=black, line width=1.0pt, mark=x, mark options={solid, black}, forget plot]
  table[row sep=crcr]{%
3.00000000000009	0.0274216\\
5.99999999999991	0.338601\\
8.99999999999989	0.259324\\
11.9999999999998	0.0431056\\
14.9999999999998	0.0270714\\
17.9999999999997	0.23579\\
20.9999999999996	0.187994\\
23.9999999999996	0.0462301\\
26.9999999999995	0.0389745\\
29.9999999999994	0.198172\\
33.0000000000008	0.190962\\
36.0000000000007	0.0667873\\
39.0000000000007	0.0667873\\
42.0000000000006	0.173211\\
};
\addplot [color=white!65!black, line width=1.0pt, mark=x, mark options={solid, white!65!black}, forget plot]
  table[row sep=crcr]{%
3.00000000000009	0.0496318\\
5.99999999999991	0.387582\\
8.99999999999989	0.188255\\
11.9999999999998	0.0433999\\
14.9999999999998	0.0395816\\
17.9999999999997	0.175575\\
20.9999999999996	0.174958\\
23.9999999999996	0.0446751\\
26.9999999999995	0.0832586\\
29.9999999999994	0.191065\\
33.0000000000008	0.223598\\
36.0000000000007	0.0456708\\
39.0000000000007	0.0454965\\
42.0000000000006	0.150026\\
};
\addplot [color=blue, line width=1.0pt, mark=x, mark options={solid, blue}, forget plot]
  table[row sep=crcr]{%
3.00000000000009	0.0274037\\
5.99999999999991	0.359835\\
8.99999999999989	0.132175\\
11.9999999999998	0.0577769\\
14.9999999999998	0.04123\\
17.9999999999997	0.102038\\
20.9999999999996	0.114905\\
23.9999999999996	0.0527644\\
26.9999999999995	0.095742\\
29.9999999999994	0.134997\\
33.0000000000008	0.146356\\
36.0000000000007	0.100402\\
39.0000000000007	0.0382314\\
42.0000000000006	0.112108\\
};
\addplot [color=red, line width=1.0pt, mark=x, mark options={solid, red}, forget plot]
  table[row sep=crcr]{%
3.00000000000009	0.0170335\\
5.99999999999991	0.234691\\
8.99999999999989	0.259852\\
11.9999999999998	0.0290367\\
14.9999999999998	0.0395876\\
17.9999999999997	0.255041\\
20.9999999999996	0.262689\\
23.9999999999996	0.061495\\
26.9999999999995	0.0198834\\
29.9999999999994	0.265786\\
33.0000000000008	0.264682\\
36.0000000000007	0.08089\\
39.0000000000007	0.0368077\\
42.0000000000006	0.137923\\
};
\end{axis}

\begin{axis}[%
	width=8.4cm,
	height=3.9cm,
	at={(0cm,0cm)},
	scale only axis,
	xmin=0,
	xmax=1,
	ymin=0,
	ymax=1,
	axis line style={draw=none},
	ticks=none,
	axis x line*=bottom,
	axis y line*=left
	]
\end{axis}
\end{tikzpicture}%